\documentclass{svmult}

\usepackage{setspace}
\usepackage{caption}
\usepackage{subcaption}
\usepackage{cite}
\usepackage{amsmath,amssymb,amsfonts}
\usepackage{graphicx}
\usepackage{textcomp}
\usepackage{xcolor}
\usepackage{lipsum}
\usepackage{arydshln}
\usepackage{mwe}
\usepackage[justification=centering]{caption}
\def\BibTeX{{\rm B\kern-.05em{\sc i\kern-.025em b}\kern-.08em
    T\kern-.1667em\lower.7ex\hbox{E}\kern-.125emX}}
    
\usepackage{lipsum}
\usepackage{mathtools}
\usepackage{cuted}
\usepackage{dsfont}
\usepackage[linesnumbered,ruled]{algorithm2e} 

\usepackage{tikz}
\usetikzlibrary{patterns}

\allowdisplaybreaks

\newcommand{\ip}[1]{\left\langle #1 \right\rangle}
\newcommand{\spn}[1]{\mbox{span}\left(#1\right) }

\definecolor{darkgreen}{HTML}{008000}
\definecolor{darkred}{HTML}{A52A2A}

\newcommand{\nvar}[2]{%
    \newlength{#1}
    \setlength{#1}{#2}
}

\nvar{\dg}{0.3cm}

\nvar{\ddx}{1.5cm}

\def\link{\draw [double distance=1.5mm, very thick] (0,0)--}
\def\joint{%
    \filldraw [fill=white] (0,0) circle (5pt);
    \fill[black] circle (2pt);
}
\def\grip{%
    \draw[ultra thick](0cm,\dg)--(0cm,-\dg);
    \fill (0cm, 0.5\dg)+(0cm,1.5pt) -- +(0.6\dg,0cm) -- +(0pt,-1.5pt);
    \fill (0cm, -0.5\dg)+(0cm,1.5pt) -- +(0.6\dg,0cm) -- +(0pt,-1.5pt);
}
\def\robotbase{%
    \filldraw[pattern=north east lines] (0,0) circle (9pt);
}

\newcommand{\angann}[2]{%
    \begin{scope}[red!40!black]
    \draw [dashed, red!40!black] (0,0) -- (1.2\ddx,0pt);
    \draw [->, shorten >=3.5pt] (\ddx,0pt) arc (0:#1:\ddx);
    \node at (#1/2-2:\ddx+8pt) {#2};
    \end{scope}
}

\newcommand{\lineann}[4][0.5]{%
    \begin{scope}[rotate=#2, blue!50!black,inner sep=2pt]
        \draw[dashed, blue!40] (0,0) -- +(0,#1)
            node [coordinate, near end] (a) {};
        \draw[dashed, blue!40] (#3,0) -- +(0,#1)
            node [coordinate, near end] (b) {};
        \draw[|<->|] (a) -- node[fill=white] {#4} (b);
    \end{scope}
}

\newcommand{\drawcm}[3]{
        \filldraw[white] (#1,#2) circle (#3);%
        \draw (#1,#2) circle (#3);%
        \fill[black] (#1,#2) -- ++(#3,0) arc (0:90:#3) -- ++(0,-2*#3) arc (270:180:#3);%
}

\def\thetaone{30}
\def\Lone{2}
\def\thetatwo{30}
\def\Ltwo{2}

\begin{document}

\title{{\LARGE \bf Regularizing Numerical Extremals Along Singular Arcs: A Lie-Theoretic Approach\thanks{This work was partially supported by AFOSR Grant FA9550-21-1-0289 and ONR Grant N00014-21-1-2431.}
}
}

\titlerunning{Lie-Theoretic Regularization of Singular Arcs}

\author{\normalsize Arthur Castello B. de Oliveira\inst{1}, Milad Siami\inst{1}, and Eduardo D. Sontag \inst{1},\inst{2}}

\institute{Department of Electrical and Computer Engineering,
Northeastern University, Boston, MA 02115 USA
	\texttt{e-mails: {\tt\small \{castello.a, m.siami, e.sontag\}@northeastern.edu}.}
 \and Department of Bioengineering,
Northeastern University, Boston, MA 02115 USA.}

\authorrunning{A.C.B. de Oliveira, M. Siami, E. D. Sontag}

\maketitle

\begin{abstract}
    Numerical ``direct'' approaches to time-optimal control often fail to find solutions that are singular in the sense of the Pontryagin Maximum Principle, performing better when searching for saturated (bang-bang) solutions. In previous work by one of the authors, singular solutions were shown to exist for the time-optimal control problem for fully actuated mechanical systems under hard torque constraints. Explicit formulas, based on a Lie theoretic analysis of the problem, were given for singular segments of trajectories, but the global structure of solutions remains unknown. In this work, we review the aforementioned framework, and show how to effectively combine these formulas with the use of general-purpose optimal control software packages. By using the explicit formula given by the theory in the intervals where the numerical solution enters a singular arc, we not only obtain an algebraic expression for the control in that interval but we are also able to remove artifacts present in the numerical solution. In this way, the best features of numerical algorithms and theory complement each other and provide a better picture of the global optimal structure. We illustrate the technique on a two degree of freedom robotic arm example, using two distinct optimal control numerical software packages running on different programming languages.
\end{abstract}

\section{Introduction}
\label{sec:Intro}

Traditionally, one of the key problems in the control of robotic systems is that of path planning, as evident by the multiple techniques formulated in the literature. 
Among those methods, collocation-based approaches show good results for solving optimal control problems both in simulation and practical applications \cite{patterson2014gpops,hager2016convergence,von1993numerical,elnagar1997collocation,pager2022method}, providing approximately optimal trajectories to follow. Alternatively, kinodynamic path planning approaches\cite{canny1988complexity,donald1993kinodynamic,hsu2002randomized} have a large literature on approximate or case-specific solutions for finding optimal trajectories \cite{canny1988complexity,webb2013kinodynamic,karaman2010optimal,donald1993kinodynamic,hsu2002randomized}. 
     
While the class of optimal control problems is very diverse, a very important type of problem, and of particular interest in this paper, is the minimum time point-to-point control problem. Its interest in the field of robotics is illustrated, for example, in disaster scenarios, where it is essential to complete tasks as rapidly as possible, or in industrial applications, where it means not only making the most out of the invested parts, but also a direct increase in production and efficiency.
The interest in this type of optimal control problem is such that there is much work in the literature \cite{bobrow1988optimal,time1992number} to cast the time-optimal control problem in a solvable way for numerical algorithms in general.

Complementary to optimal path planning, optimal trajectory tracking is also important to guarantee that the resulting control law is optimal, or near optimal, with works in the literature also focusing on the time-optimal aspect of this problem. For example, in \cite{bobrow1985time} the authors consider the time-optimal control of manipulators across a pre-specified trajectory, looking at the maximum rate at which one can traverse it while still keeping the trajectory admissible. 

However, despite the vast literature on both path planning and trajectory tracking, numerical methods for finding an optimal trajectory, can struggle to recover the control signal whenever the switching function becomes singular, also known as solutions with singularity arcs. In \cite{pager2022method,pager2022computational} the authors develop a way of identifying regions of the solution with singularity arcs and refining the resulting controller from their collocation method, however in some cases, numerical artifacts are still present on the recovered solution. 

This paper explores a theoretical formulation for fixed start and endpoint minimum time control of manipulators in the presence of singularity arcs. The resulting controller is guaranteed to satisfy the Pontryagin Maximum Principle, which is a necessary condition for optimality and matches the trajectories obtained by off-the-shelf optimal control software while giving a closed-form solution for the control signals and avoiding numerical artifacts.

We base our method on theoretical work explored by one of the authors \cite{sontag1985remarks,sontag1986time}, and while the theoretical framework we use was already shown in these previous publications, in this work we consolidate the results, provide a different format for some of the proofs, and fix a few minor mistakes in these previous works. Regardless, this theoretical approach leverages the Lie algebraic structure of the system to guarantee that the generated trajectories satisfy the Pontryagin Maximum Principle. After presenting some simulations for a two-degree-of-freedom (2DOF) robotic arm model in which singularity arcs appear in the optimal solution, we show how to use theory to regularize the numerical solution. 

\section{Theoretical Background}

\subsection{Fully Actuated Mechanical Systems and the Minimum-Time Control Problem}

We consider what is typically called a ``fully actuated mechanical systems'' whose dynamics are given by
\begin{equation}
\label{eq:SOMechSysEq}
 u(t) \;=\; M(q(t))\,\ddot q(t)\,+\,C(q(t),\dot q(t))\,+\,G(q(t)) \,.
\end{equation}
Here the components of  $u:\mathbb{R}^+\rightarrow\mathbb{R}^n$, $u(t) = [u_1(t), ..., u_n(t)]^\top$ are seen as inputs (forces and torques)
and
the components of the vector function $q:\mathbb{R}^+\rightarrow\mathbb{R}^n$ are called the configuration variables of the system (linear and angular displacements). The ``inertia matrix'' $M(q)$ is a positive definite matrix function for every $q\in\mathcal{Q}$, $\mathcal{Q}$ being the set of admissible trajectories of the system. 
To provide a more standard state-space form, we let $x:=[q^\top, ~\dot q^\top]^\top$ be the state vector of the system, and define $L(q) := M^{-1}(q)$. Then the state-space equations of the system are given by
\begin{equation}
\label{eq:GeneralStateDynamics}
\underbrace{\begin{bmatrix}
                \dot{q} \\ \ddot{q}
            \end{bmatrix}}_{\dot x} = \underbrace{\begin{bmatrix}
                \dot q \\ -L(q)(C(q,\dot q)+G(q))
            \end{bmatrix}}_{f(x)} + \underbrace{\begin{bmatrix}
                0 \\ L(q)
            \end{bmatrix}}_{g(x)}u.
\end{equation}

The problem of interest of this paper is as follows:
given initial and final desired configurations $x_{0}$ and $x_{f}$, and bounds on the components of the control signal
\[
L_i\leq u_i(t)\leq M_i,
\]
for constant $L_i,M_i\in\mathbb{R}$ and $\forall i\in\mathbb{N}_{\leq n}$, one wishes to find control functions $u^*(\cdot)$ such that
\[
\varphi(T,x_{0},u^*) = x_{f}
\]
with the smallest possible $T$ (where $\varphi(t,\xi,u)$ is the solution at time $t$ if the initial condition is $\xi$ and the input is $u(\cdot)$, assuming that a solution is well-defined on $[0,t]$). This problem can be expressed as:
\begin{equation}
\label{eq:ContinuousTimeMinimumTimeControlProblem}
            \begin{array}{rlclcl}
            \displaystyle \min_{u\in\mathcal{M}} & T\\
            \textrm{s.t.} & \dot x(t) = f(x(t))+g(x(t))u(t)\\
            &x(0) = x_{i}   \\
            &x(T) = x_{f}  \\
            \end{array}
        \end{equation}
where $\mathcal{M}$ is the set of measurable essentially bounded functions
\[
u:[0,T]\rightarrow\mathbb{R}^n
\]
such that
$L_i\leq u_i(t)\leq M_i$
for almost all $t\in[0,T]$.

Such an optimization problem is not always analytically solvable, or even numerically solvable in a reasonable time. For this reason, numerical methods are often employed to search for approximately/near-optimal control strategies. Knowledge of the theory behind such optimization problems can prove very useful in understanding the obtained solution and even regularizing numerical artifacts that might be present, as we argue in this paper.

In the next section, we look at what a solution for the time-optimal control looks like and how we can use properties of our system to obtain algebraic expressions for our control in cases where the trajectory has singular sections.
        
\subsection{Extremals and the Maximum Principle}

Candidate solutions of optimal control problems are obtained through the Pontryagin Maximum Principle (PMP) \cite{seierstad1977sufficient,pontryagin2018mathematical}. Define the Hamiltonian associated to problem \eqref{eq:ContinuousTimeMinimumTimeControlProblem} as
        \begin{equation}
            \label{eq:uising_hamiltonian}
            H(x,u,\lambda) := \ip{\lambda,(f(x)+g(x)u)}-1
        \end{equation}
where $\lambda:\mathbb{R}_+\rightarrow\mathbb{R}^{2n}$ is called the costate of the system and follows the adjoint dynamics defined as
        \begin{equation}
            \label{eq:uising_adj}
            \dot\lambda = -\ip{\frac{\partial}{\partial x} (f(x)+g(x)u),\lambda}.
        \end{equation}
    
Let $I=[0, T]$, $T>0$, $x^*:I\rightarrow\mathbb{R}^{2n}$,
and $u^*:I\rightarrow \mathcal{M}_0$ where 
\[
\mathcal{M}_0:=\{[u_1,\dots,u_n]^\top\in\mathbb{R}^n~|~L_i\leq u_i\leq M_i, ~\forall i=1,\dots,n\}
\]
and such that $(x^*,u^*)$ satisfy \eqref{eq:GeneralStateDynamics}. The Pontryagin Maximum Principle, then, states that if $(x^*,u^*)$ solves the minimum-time control problem for $x_0=x^*(0)$ and $x_f=x^*(T)$, then there exist some $\lambda^*:I\rightarrow\mathbb{R}^{2n}$, $\lambda^*(t)\neq 0$, that satisfies \eqref{eq:uising_adj}, such that $(x^*,u^*,\lambda^*)$ satisfies
        \begin{equation}
            \label{eq:StatementMaximumPrinciple}
            H(x^*,u^*,\lambda^*)=\max_{a\in\mathcal{M}_0}H(x^*,a,\lambda^*).
        \end{equation}
(Observe that asking $\lambda^*(\bar t)\not=0$ for some $\bar t$ is equivalent to asking $\lambda^*(t)\not=0$ for all $t\leq \bar t$, because the equation for $\lambda^*$ is linear.)
Any tuple $(x^*, u^*, \lambda^*)$ that satisfies the maximum principle is called an \emph{extremal} of the system. Furthermore, notice that because of the particular structure of our system, we can define 
        \begin{equation}
            \frac{\partial H}{\partial u_i}(x(t),\lambda(t)) = \ip{\lambda(t), g_i(x(t))} = \phi_i(t) 
        \end{equation}
and conclude from \eqref{eq:StatementMaximumPrinciple} that $u_i(t)=M_i$ if $\phi_i(t)>0$ and $u_i(t)=L_i$ if $\phi(t)<0$. The function $\phi_i$ is commonly called the \emph{switching function} associated with the input $u_i$. If a switching function has a finite number of zeros in $I$ then the extremal associated to it is called \emph{$u_i$-bang-bang extremal}, and $u_i$ is piecewise constant. If, however, there is a non-trivial sub-interval $\bar I$ of $I$ in which $\phi_i\equiv 0$, then  the PMP maximization condition~\eqref{eq:StatementMaximumPrinciple} is not sufficient to characterize the behavior of $u_i$ in $I$, and the extremal is called \emph{$u_i$-singular}.

\subsection{On Singular Extremals and Their Existence}

Consider an extremal $(x,u,\lambda)$ and, for all $i$ between $1$ and $n$, let $J_i$ be the set of points in $I$ such that $\phi_i=0$ and $\phi_i'=0$. The time derivative of the switching function can be computed almost everywhere as
        \begin{align*}
            \phi_i' &= \ip{\dot\lambda(t),g_i(x(t))}+\ip{\lambda(t),\frac{\partial g_i(x)}{\partial x}\dot x}
            \\ &= \ip{\lambda(t), [f,g_i](x(t))+\sum_{j=1}^nu_j(t)[g_j,g_i](x(t))},
        \end{align*}
where $[\cdot,\cdot]$ is the Lie Bracket of two vector fields defined as $[a(x),b(x)] = \frac{\partial b}{\partial x}a-\frac{\partial a}{\partial x}b$. The vector-fields $g_i$ are given by the columns of $g = [0, L(q)^\top]^\top$ which means that the $g_i$s have the structure of $g_i = [0, \ell_i(q)^\top]^\top$, where $\ell_i$ is the $i$-th column of $L$. From this we can compute
\begin{equation*}
            \frac{\partial g_i}{\partial x} = \begin{bmatrix}
                0 & 0 \\ \frac{\partial \ell_i(q)}{\partial q} & 0
            \end{bmatrix}
        \end{equation*}
which in turn implies that for any $i,j$, $[g_i,g_j]=0$. 
From this, it follows that
\begin{align*}
\phi_i' &= \ip{\lambda(t), [f,g_i](x(t))}
        \end{align*}
and in particular the function $\phi_i'$ is again differentiable.
Similarly, one can show that 
        \begin{equation*}
            [f,g_i] = -\begin{bmatrix}
                \ell_i(q) \\ *
            \end{bmatrix}
        \end{equation*}
which in turn implies that $\{g_1,\dots,g_n,[f,g_1],\dots,[f,g_n]\}$ is a frame of vector fields. This observation derived from the assumed structure of our system allows us to state the following result regarding the existence of singular extremals:

\begin{lemma}[\cite{sontag1985remarks,sontag1986time}]
\label{lemma:Jis}
The set of points for which all switching functions and their derivatives are zero is empty, i.e. $J_1\cap J_2 \cap J_3 \dots \cap J_n= \emptyset$.
        \end{lemma}

This follows immediately from the fact that $\phi_i = \ip{\lambda,g_i}=0$ and $\phi_i'=\ip{\lambda,[f,g_i]}=0$ hold for all $i$ if and only if $\lambda\equiv 0$ (since $\{g_1,\dots,g_n,[f,g_1],\dots,[f,g_n]\}$ is a frame) contradicting the assumption that $(x,u,\lambda)$ is an extremal. From this, the following Corollary follows immediately:

\begin{corollary}
    If the extremal $(x,u,\lambda)$ is $u_i$-singular for all $i\neq k$ for some $k\in\mathbb{N}_{\leq n}$, then $u_k$ is bang-bang.
\end{corollary}

For evaluating the singularity conditions, we compute the second derivative of the $i$-th switching function as below:
        \begin{align*}
            \phi_i'' &= \ip{\dot\lambda, fg_i}+\ip{\lambda,\frac{\partial}{\partial x}(fg_i)\dot x}\\
            &= -\ip{\left(f'+\sum_{j=1}^n u_jg'_j\right)^\top\lambda,fg_i} + \ip{\lambda,\frac{\partial}{\partial x}(fg_i)\left(f+\sum_{j=1}^n u_jg_j\right)} \\
            &= \ip{\lambda, \frac{\partial}{\partial x}(fg_i)\left(f+\sum_{j=1}^n u_jg_j\right)-\left(f'+\sum_{j=1}^n u_jg'_j\right)fg_i} \\
            &= \ip{\lambda, ffg_i}+\sum_{j=1}^n u_j\ip{\lambda,g_jfg_i}
        \end{align*}
where from now on, to simplify notations, we write iterated Lie brackets as
\[
X_1X_2\ldots X_p = 
[X_1,[X_2,[\ldots,[X_{p-1},X_p]\ldots]]]\,.
\]
To further simplify this expression, consider the following proposition:
\begin{proposition}[\cite{sontag1985remarks,sontag1986time}]
\label{prop:gifgj}
For mechanical systems with dynamics given by \eqref{eq:SOMechSysEq}, for any $i,j\in[1,~2,~\dots,~n]$, $g_ifg_j\in\mbox{span}(\{g_1,\dots,g_n\})$. That is, $\exists \alpha_{ijk}:\mathbb{R}^{2n}\rightarrow\mathbb{R}$ such that
\begin{equation}
                (g_ifg_j)(q,\dot q) = \sum_k\alpha_{ijk}([q;\dot q])g_k(q).
            \end{equation}
        \end{proposition}

One can easily verify this algebraically: remembering that $g_i=[0,~\ell_i(q)^\top]^\top$ and $fg_j = -[\ell_j(q)^\top, ~*]^\top$, then 
        \begin{align}
            g_ifg_j &= \frac{\partial fg_j}{\partial x}^\top g_i-\frac{\partial g_i}{\partial x}^\top fg_j \nonumber\\
            &= -\begin{bmatrix}
                \frac{\partial \ell_i^\top}{\partial q} & 0 \\ * & *
            \end{bmatrix}\begin{bmatrix}
                0 \\ \ell_j(q)
            \end{bmatrix}+\begin{bmatrix}
                0 & 0 \\ * & *
            \end{bmatrix}\begin{bmatrix}
                \ell_j \\ *
            \end{bmatrix} \nonumber\\ &= \begin{bmatrix}
                0 \\ *
            \end{bmatrix}.
        \end{align}
        With this, we can write the second derivative of the switching function as
        \begin{align}
            \phi_i'' &= \ip{\lambda, ffg_i+\sum_{j=1}^n u_j\sum_{k=1}^n \alpha_{ijk}g_k}\nonumber \\
            &= \ip{\lambda, ffg_i+\sum_{k=1}^n \sum_{j=1}^n u_j\alpha_{ijk}g_k} \\ &= \ip{\lambda, ffg_i+\sum_{k=1}^n \beta_{ik}g_k} \nonumber \\ &= \ip{\lambda,ffg_i}+\sum_{k=1}^n \beta_{ik}\phi_k. \label{eq:d2phi}
        \end{align}
        where all time dependencies were omitted for clarity, also, notice that $\beta_{ik}$ are also a function of the control signals.

Notice that expression \eqref{eq:d2phi} holds in general, without any assumption of singularity.
Next, for some $k\in\mathbb{N}_{\leq n}$ assume the system is $u_i$-singular for all $i\in\mathbb{N}_{\leq n}$, $i\neq k$, then
        \begin{equation}
            \label{eq:sndodrcnd}
            \phi_i(t)'' = \ip{\lambda, ffg_i}+\beta_{ik}\phi_k=0.
        \end{equation}
Let us consider the following set of vector fields:
        \begin{equation*}
            \{g_i~|~i\in\mathbb{N}_{\leq n}\}\cup\{fg_i~|~i\in\mathbb{N}_{\leq n},~i\neq k\}\cup\{ffg_i ~|~i\in\mathbb{N}_{\leq n},~i\neq k\},
        \end{equation*}
and let $S_k$ be the set of states for which these vector fields span the entire tangent space. From this, {the following Theorem can be proven by contradiction}:
        
\begin{theorem}[\cite{sontag1985remarks,sontag1986time}]
If the extremal is $u_i$-singular for all $i\in\mathbb{N}_{\leq n},~i\neq k$ and is inside $S_k$ for all $t$, then $u_k$ is constant and equal to $L_k$ or constant equal to $M_k$ almost everywhere (i.e. for all time $t$, except for a set of measure zero), i.e. there is no switching of the value of $u_k$.
\end{theorem}

Let $u_k=c_k$ where $c_k$ is constant and equal to either $M_k$ or $L_k$. Let $\bar u$ be the vector obtained by concatenating all $u_i$ for $i\neq k$. To determine the expression for the singular controls, expand the second-order singularity condition as follows:

\begin{align*}
    0&=\phi_i(t)''\\
    &= \ip{\lambda,ffg_i}+c_k\alpha_{ikk}\phi_k+\sum_{j\neq k}u_j\alpha_{ijk}\phi_k \\
    &= \ip{\lambda,ffg_i}+c_k\alpha_{ikk}\phi_k+ \bar u^\top a_{ik}\phi_k
\end{align*}
where $a_{ik}$ is the column vector obtained from concatenating $\alpha_{ijk}$ for $j\neq k$. Writing the equation above for all $i\neq k$ and concatenating it in matrix form results in

\begin{equation}
    \label{eq:uisolanalyt}
    0 = \psi_k + b_{kk}c_k\phi_k+\bar u^\top A_k\phi_k
\end{equation}
where $\psi_k$ is the vector obtaining from concatenating $\ip{\lambda,ffg_i}$ for $i\neq k$, $b_{kk}$ is the vector obtained from concatenating $\alpha_{ikk}$ for $i\neq k$ and $A_k$ is the matrix whose columns are given by $a_{ik}$ for $i\neq k$. One can easily verify that if $\Delta_k=\det(A_k)\neq 0$, and $\phi_k\neq 0$, then there is a unique solution for $\bar u$ from solving \eqref{eq:uisolanalyt}. With this, define:

\begin{equation*}
    R_k = S_k\cap\{x~|~\Delta_k(x)\neq 0\}
\end{equation*}
then we state the following Theorem:

\begin{theorem}[\cite{sontag1985remarks,sontag1986time}]
    \label{thm:ukanaly}
        Given an extremal of the time-optimal control problem, if it is $u_i$ singular for all $i\neq k$ and remains in $R_k$ for all $t$, then all $u_i$ are analytic functions of time and can be computed by solving \eqref{eq:uisolanalyt}.
    \end{theorem}

        With this we have set up the theoretical background to regularize numerical artifacts in singular solutions obtained from numerical solvers for optimal control problems.
        
    \section{Time-Optimal Control of a Robotic Arm}
\label{sec:uising}
    In this section, we illustrate the discussed technique on a 2-DOF robotic arm. This system provides a good example of the theory since its reduced order together with Lemma \ref{lemma:Jis} implies that only one of the controllers can be singular at any given time. Furthermore, as we will see shortly, the second actuator can never be singular by itself, implying that any trajectory with a singular arc is such that $\phi_1(t)\equiv 0$ for some non-trivial time interval.
    
    \subsection{The $2$-DOF Arm Model}
        
        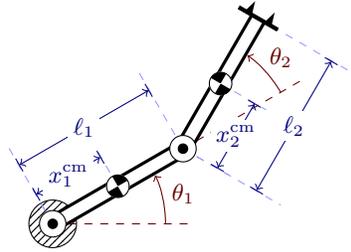
\begin{figure}
            \centering
            \begin{tikzpicture}
                \robotbase
                \angann{\thetaone}{\footnotesize$\theta_1$}
                \lineann[1.2]{\thetaone}{\Lone}{\footnotesize$\ell_1$}
                \lineann[0.6]{\thetaone}{\Lone/2}{\footnotesize$x^{\text{cm}}_1$}
                \link(\thetaone:\Lone);
                \joint
                \begin{scope}[shift=(\thetaone:\Lone), rotate=\thetaone]
                    \drawcm{-\Lone/2}{0}{0.15}
                    \angann{\thetatwo}{\footnotesize$\theta_2$}
                    \lineann[-1.5]{\thetatwo}{\Ltwo}{\footnotesize$\ell_2$}
                    \lineann[-0.7]{\thetatwo}{\Ltwo/2}{\footnotesize$x^{\text{cm}}_2$}
                    \link(\thetatwo:\Ltwo);
                    \joint
                    \begin{scope}[shift=(\thetatwo:\Ltwo), rotate=\thetatwo]
                        \drawcm{-\Ltwo/2}{0}{0.15}
                        \grip
                    \end{scope}
                \end{scope}
            \end{tikzpicture}
            \caption{\small Diagram of a 2 link planar robotic manipulator. Gravity is assumed to be orthogonal to the plane of movement of the robot. All indicated parameters are given in Table \ref{tab:arm2dof_param}. }
            \label{fig:arm2dof}
        \end{figure}
    
        \begin{table}[]
            \centering
            \begin{tabular}{|c|c|c|}
                \hline
                Link & \#1 & \#2 \\ \hline \hline
                Length ($\ell$) & 0.5 & 0.5\\ \hline
                CoM Position ($x^{\text{cm}}$) & 0.5 & 0.5\\ \hline
                Mass ($m$) & 50 & 30\\ \hline
                Inertia ($I_z$) & 5 & 3 \\ \hline
            \end{tabular}
            \caption{\small Parameters for the 2DOF manipulator depicted in Fig. \ref{fig:arm2dof}.}
            \label{tab:arm2dof_param}
        \end{table}
        
        Consider the 2-link manipulator illustrated in Fig. \ref{fig:arm2dof}, with parameters given in Table \ref{tab:arm2dof_param}. Assuming that gravity is perpendicular to the plane of movement results in the following equations of motion:
        \begin{equation}
            \label{eq:arm2dof_EoM}
            u ~=~ M(\theta)\ddot{\theta}+C(\theta,\dot{\theta})
        \end{equation}
        where
        \begin{equation*}
            \begin{split}
                M_{11} ~=~ &m_2\ell_1^2+2m_2\cos(\theta_2)\ell_1 x^{\text{cm}}_{2}+m_1\left(x^{\text{cm}}_{1}\right)^2\\&+m_2\left(x^{\text{cm}}_{2}\right)^2+I_{z,1}+I_{z,2} \\
                M_{12}~=~ &M_{21}=m_2\left(x^{\text{cm}}_{2}\right)^2+\ell_1m_2\cos(\theta_2)x^{\text{cm}}_{2} \\
                M_{22} ~=~ &m_2\left(x^{\text{cm}}_{2}\right)^2+I_{z,2} \\
                C_1 ~=~ &-\ell_1m_2x^{\text{cm}}_{2}\sin(\theta_2)\dot\theta_2^2\\&-2\ell_1\dot\theta_1 m_2x^{\text{cm}}_{2}\sin(\theta_2)\dot\theta_2 \\ C_2 ~=~ &\ell_1m_2x^{\text{cm}}_{2}\sin(\theta_2)\dot\theta_1^2 \\
                \theta ~=~ &[\theta_1, ~\theta_2]^\top \\
                u ~=~ &[u_1, ~u_2]^\top.
            \end{split}
        \end{equation*}
    
        The state-space equations are, then, given by
    
        \begin{equation}
            \label{eq:arm2dof_sseq}
            \underbrace{\begin{bmatrix}
                \dot{\theta} \\ \ddot{\theta}
            \end{bmatrix}}_{\dot x} = \underbrace{\begin{bmatrix}
                \dot\theta \\ -M(\theta)^{-1}C(\theta,\dot\theta)
            \end{bmatrix}}_{f(x)} + \underbrace{\begin{bmatrix}
                0 \\ M(\theta)^{-1}
            \end{bmatrix}}_{g(x)}u.
        \end{equation}  
    
        This model, albeit simple, serves as a basis for comparison and validation between different time-optimal control strategies. Next we take a look at what the singular solutions look like for this model.

    \subsection{Singular Solutions for the Arm Model}

        The 2DOF arm model is significantly simpler than the general case dealt in the previous section since it possesses only two actuators. As a consequence, the possible singular controls can be analyzed by exhaustion.

        \subsubsection{On the $u_2$-Singularity}

            For the 2-DOF arm case, symbolic computation shows that $\alpha_{ij1}\equiv 0$, which in turn implies that $\Delta_{1} = 0$. This means that \eqref{eq:sndodrcnd} is simplfied to

            \begin{equation*}
                \phi_2'' = \ip{\lambda,ffg_2} = 0.
            \end{equation*}

            We, then, compute the third-order condition for singularity, given by

            \begin{equation}
                \phi_2''' = \ip{\lambda, fffg_2+u_1g_1ffg_2+u_2g_2ffg_2}.
            \end{equation}

            Again, for the 2-DOF arm case, it can be shown symbolically that $\ip{\lambda,g_2ffg_2}$ is a linear combination of $\phi_2$ and $\phi_2'$, and therefore vanishes by singularity. Furthermore, $u_1$ must be constant equal to $L_1$ or $U_1$. Let $B$ be the set where the vectors $\{g_2, fg_2, ffg_2, fffg_2+cg_1ffg_2\}$ are linearly independent for $c=L_1$ or $c=U_1$. The set $B$ is open and can be shown to be nonempty; thus we can state:

            \begin{theorem}[\cite{sontag1985remarks,sontag1986time}]
                There are no $u_2$-singular extremals for which the state $x(t)$ intersects the open dense set $B$.
            \end{theorem}

            From this, we next focus on the case where the first control signal exhibits singularity.

        \subsubsection{On the $u_1$-Singularity}

            For this section, we assume $k=2$, that is, the extremal is $u_1$-singular and (consequently) $u_2$-bang-bang. We can verify that along an $u_1$-singular extremal of the 2DOF arm system, it is necessary that:
            \begin{equation}
                \label{eq:uising_swtsing}
                \begin{split}
                    \phi_1 &= \ip{\lambda,g_1} = 0 \\
                    \phi_1' &= \ip{\lambda,fg_1} = 0 \\
                    {\phi_1''} &= \ip{\lambda,ffg_1}+(\alpha_{1}(x) u_1+\alpha_2(x)u_2) \ip{\lambda,g_2} = 0.
                \end{split}
            \end{equation}
            
            The functions $\alpha_{1,2}$ are not simultaneously zero, and can be obtained by solving:
            \begin{equation}
                g_1fg_2 = \alpha_{1}g_1+\alpha_2g_2
            \end{equation}
            which always has a solution since $g_1fg_2\in\spn{\{g_1,g_2\}}$, as per Proposition \ref{prop:gifgj}.
        
            Next, notice that $g_1 = [0 ; 0 ; \mu ; \nu]$ and $fg_1 = [-\mu ; -\nu ; 0 ; \gamma]$, where $\mu(x), \nu(x)$ and $\gamma(x)$ are scalar functions of the states. Then, the first two equations of \eqref{eq:uising_swtsing}, which give us orthogonality conditions for $\lambda$, are equivalent to:
            \begin{equation}
                \label{eq:uising_ldecomp}
                \lambda(t) = \lambda_2(t)\underbrace{\begin{bmatrix}-\nu/\mu \\ 1 \\ 0 \\ 0\end{bmatrix}}_{a(x)}+\lambda_4(t)\underbrace{\begin{bmatrix}\gamma/\mu \\ 0 \\ -\nu/\mu \\ 1\end{bmatrix}}_{b(x)}\,.
            \end{equation}
        
            As a consequence of decomposing $\lambda(t)$ as in \eqref{eq:uising_ldecomp}, $\ip{a(x),g_2}=0$, which together with Lemma \ref{lemma:Jis}, means we can rewrite the third equation in \eqref{eq:uising_swtsing} as:
            \begin{equation}
                \label{eq:uising_u1cont}
                u_1(x,\lambda) = r(x)\frac{\lambda_2}{\lambda_4}+s(x)
            \end{equation}
            where, since $u_2=c$ where $c= M_2$ or $c=L_2$:
            \begin{equation*}
                \begin{split}
                    r(x) &= -\frac{1}{\alpha_1(x)}\frac{\ip{a(x),ffg_1(x)}}{\ip{b(x),g_1(x)}} \\
                    s(x) &= -\frac{1}{\alpha_1(x)}\frac{\ip{b(x),ffg_1(x)}}{\ip{b(x),g_1(x)}}-\frac{\alpha_2(x)}{\alpha_1(x)}c.
                \end{split}
            \end{equation*}
        
            Therefore, we can recover an expression for the control signal $u_1$ that maintains $u_1$-singularity as long as $\lambda_4(t)\neq 0$ and $\alpha_1(x)\neq 0$, since if either of those fail, then our control expression is undefined. To deal with this problem we restrict our trajectories to a subspace of our state space defined as
            \begin{equation}
                \small
                \begin{split}
                    R_k = \{x\in\mathbb{R}^4 ~&|~ \spn{\{g_1(x),g_2(x),(fg_1)(x),(ffg_1)(x)\}}=\mathbb{R}^4, ~\\&\text{and}~\alpha_1(x)\neq 0\}.
                \end{split}
            \end{equation}
        
            One can verify that if $x(t)\in S_k$ then $\lambda_4(t)\neq 0$ as per Theorem \ref{thm:ukanaly}. Furthermore, the invertibility of $\alpha_1(x)$ is necessary for $r(x)$ and $s(x)$ to be well defined.
        
            For our 2DOF arm, we can explicitly write $R_k$ after evaluating our symbolic expressions as
            \begin{equation}
                \label{eq:uising_R}
                R_k = \{x\in\mathbb{R}^4~|~\theta_2\neq k\pi/2, ~\text{and}~\dot\theta_1+\dot\theta_2\neq0\}
            \end{equation}
            and as long as our trajectory remains in this set we can choose $u_1$ as in \eqref{eq:uising_u1cont} and enforce $\phi_1\equiv 0$. The next session describes a process through which a $u_1$-singular extremal can be analytically found for the 2DOF arm system. After that, numerical tools are used to attempt to recover the obtained singular extremal, or a more efficient solution for the same initial and endpoints.

    \subsection{Numerical Simulations}

        We first obtain a $u_1$-singular extremal of the 2DOF arm by numerically solving \eqref{eq:GeneralStateDynamics} and \eqref{eq:uising_adj} for a fixed time $T=0.7$, and imposing the following initial conditions on the states and costates:
        \begin{equation}
            \label{eq:arm2dof_siminitcond}
            \begin{split}
                x_0 = \begin{bmatrix}
                    \theta_0 \\ \dot\theta_0
                \end{bmatrix} = \begin{bmatrix}
                    \frac{\pi}{20} \\ \frac{\pi}{20} \\ 0.30 \\ 0.5
                \end{bmatrix},~~
                \lambda_0 = \begin{bmatrix}
                      5.1165 \\ 3.0000 \\ 10.2330 \\ 6.0000
                \end{bmatrix}.
            \end{split}
        \end{equation}

        The control signal $u_2(t)=-10$ is chosen to be saturated at its lower-bound, and $u_1$ is computed as in \eqref{eq:uising_u1cont} and has as saturation $|u_1|\leq 20$. The resulting trajectory is guaranteed to be an extremal of the system and is used as a baseline to compare with numerical solvers. 

\begin{figure}
            \centering
            \includegraphics[scale=0.63]{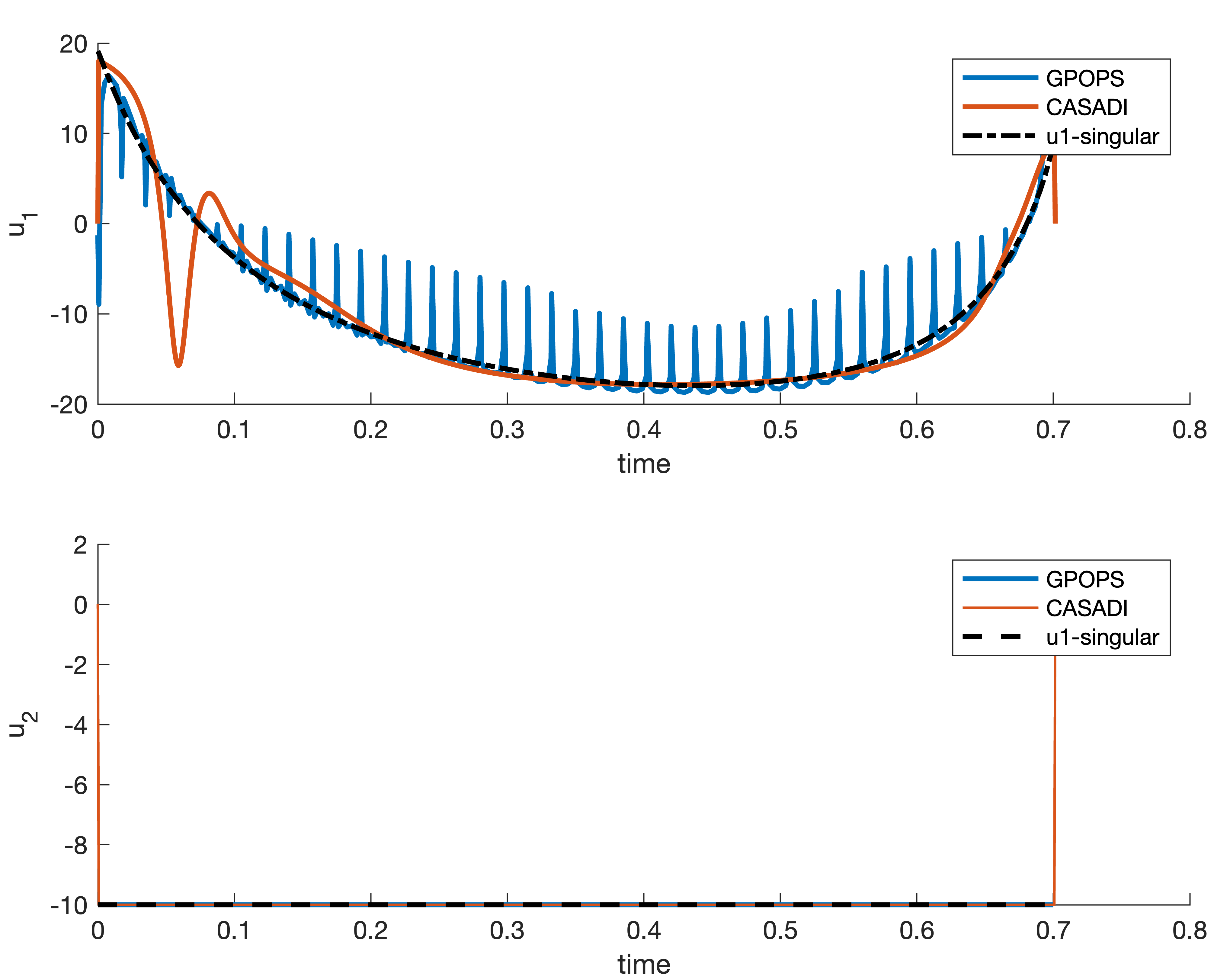}
            \includegraphics[scale=0.63]{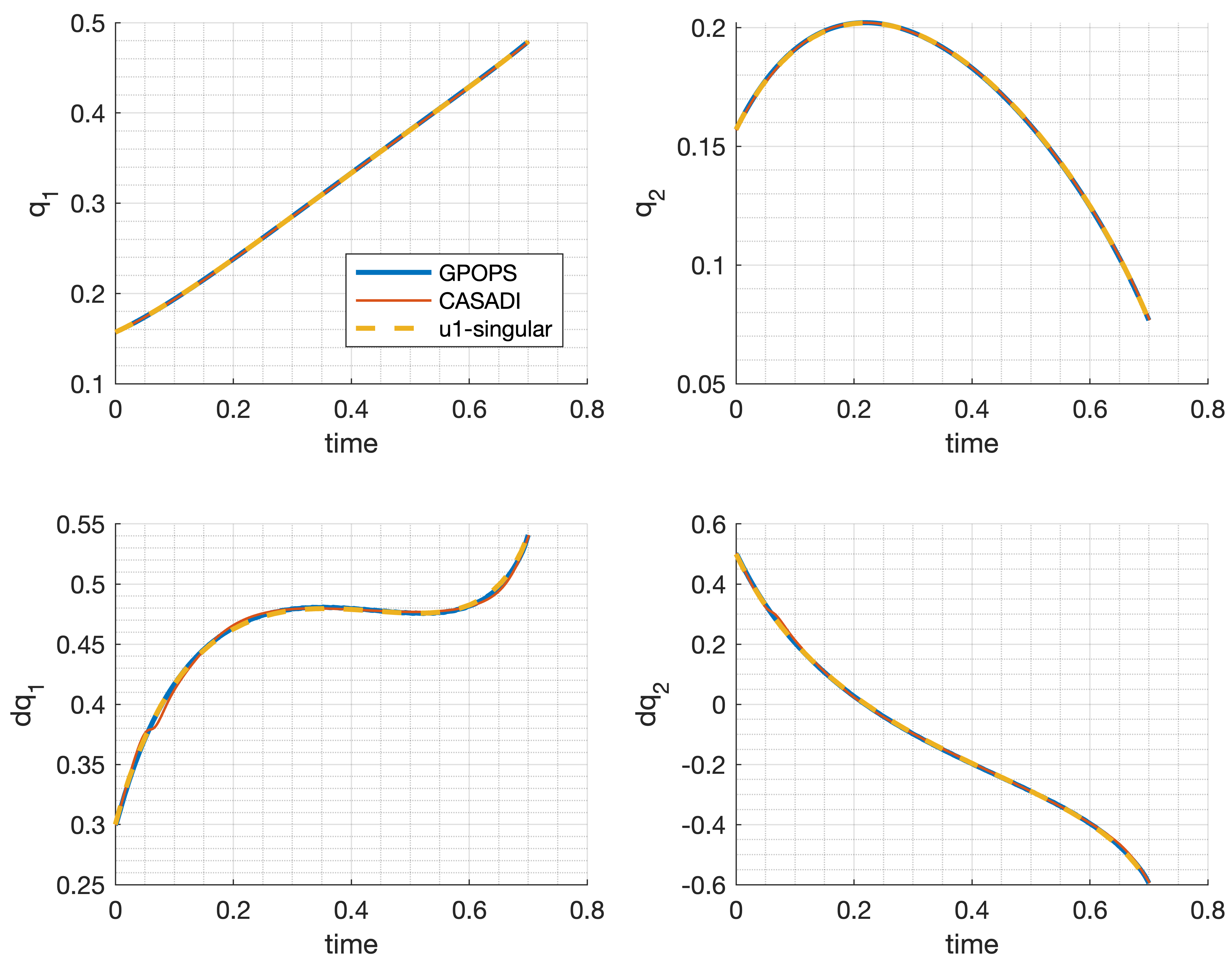}
            \caption{\small Comparison between a $u_1$-singular extremal (dashed black) and the numerical solutions obtained by GPOPS-II (solid blue) and CASADI (solid red). All solutions reach the target state at basically the same time, and follow
            the same trajectory for the states. Furthermore, they appear to follow the same profile for the input signals, albeit presenting numerical artifacts in the case of the numerical solutions.}
            \label{fig:arm2dof_u1singsimul}
        \end{figure}

        We next give the same initial condition $x_0$ and the resulting endpoint $x(T)$ (from solving the above equation) to GPOPS-II \cite{patterson2014gpops}, an off-the-shelf, general-purpose optimal control software, and CASADI, an auxiliary toolbox that facilitates the solving of optimal control problems. We ask them to find the minimum-time control that takes our system from $x_0$ to $x(T)$. The GPOPS-II software uses collocation in the Matlab environment to numerically find a solution that minimizes our cost function, while we built a shooting-based code with the help of CASADI in Python. This means that both software packages approach the problem differently from each other and from our constructive control based on the Pontryagin Maximum Principle and singularity arcs. The resulting trajectories are plotted on top of the ones obtained from $u_1$-singularity and presented in Fig. \ref{fig:arm2dof_u1singsimul}.
        
        From Fig. \ref{fig:arm2dof_u1singsimul} we can see that both GPOPS-II and CASADI reach the same optimal-time value as the $u_1$-singular control, and both attempt to reconstruct \eqref{eq:uising_u1cont}, albeit with numerical artifacts. While none of the solutions are guaranteed to be the global minimum of the problem, the fact that they reached the same cost and trajectory through different methodologies gives us confidence in the results. Furthermore, the state trajectories for all three controls are the same.

\begin{figure}
            \centering
            \includegraphics[scale=1]{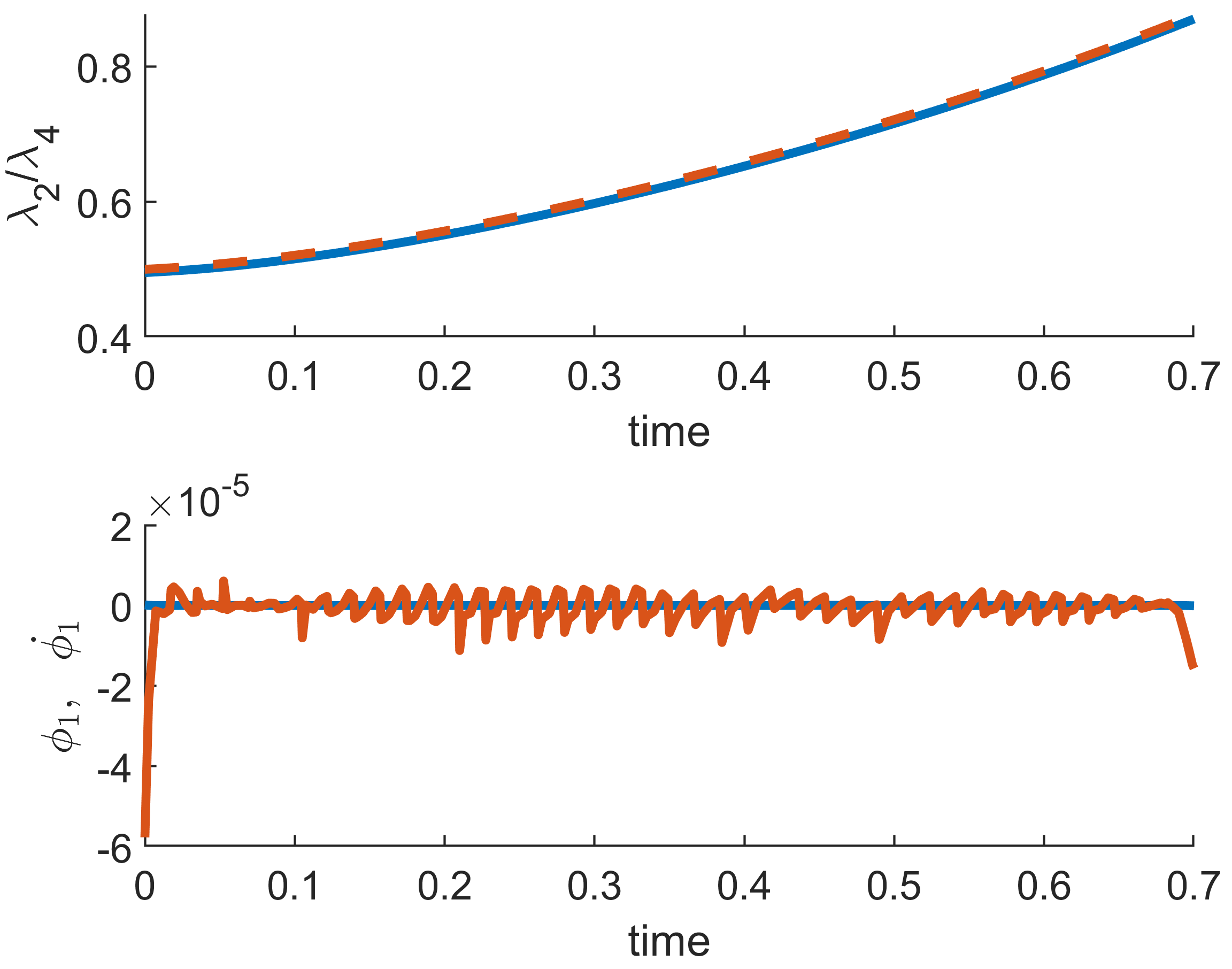}
            \caption{\small In this figure we can see in the top plot that GPOPS-II (solid blue) recovers the same ratio $\lambda_2/\lambda_4$ than our $u_1$-singular extremal (dashed red). Furthermore, on the bottom plot we see $\phi_1=\ip{\lambda,g_1}$ (blue) and $\phi_1'=\ip{\lambda,fg_1}$ (red). We can see that both signals are very close to zero, but $\dot\phi_1$ present very small oscillations which look to correlate to the peaks present in the signal for $u_1$ in Fig. \ref{fig:arm2dof_u1singsimul}.}
            \label{fig:arm2dof_gpopsvsu1sing}
        \end{figure}

        We can deepen our analysis by examining the numerical costates returned by GPOPS-II. From Fig. \ref{fig:arm2dof_gpopsvsu1sing} we can see that both GPOPS-II and the $u_1$-singular extremal have the same value for the ratio $\lambda_2/\lambda_4$ which is the only dependency of $u_1$ on the costates in \eqref{eq:uising_u1cont}. Furthermore, also from Fig. \ref{fig:arm2dof_gpopsvsu1sing} on the bottom plot we can see that the values for both $\phi_1$ and $\dot\phi_1$ recovered by GPOPS-II are extremely small and oscillate around zero. This observation, along with the fact that to satisfy the Pontryagin Maximum Principle, $u_1$ must either be saturated at $\pm 20$ (bang-bang) or follow the expression given by \eqref{eq:uising_u1cont} (singular), allow us to definitely conclude that the peaks on the GPOPS-II solution in Fig. \ref{fig:arm2dof_u1singsimul} are indeed numerical artifacts and not part of the intended solution.

        Furthermore, we can conclude that the behavior presented by GPOPS-II when computing singular arcs is expected. Looking at the profile of $\dot\phi_1$ in Fig. \ref{fig:arm2dof_gpopsvsu1sing} shows that the algorithm struggles to achieve a value for $\phi_1$ that is identically zero, oscillating, instead, around zero with very small amplitudes. Depending on the software this might incorrectly register as a switch for the control signal.

\begin{figure}
            \centering
            \includegraphics[scale=0.6]{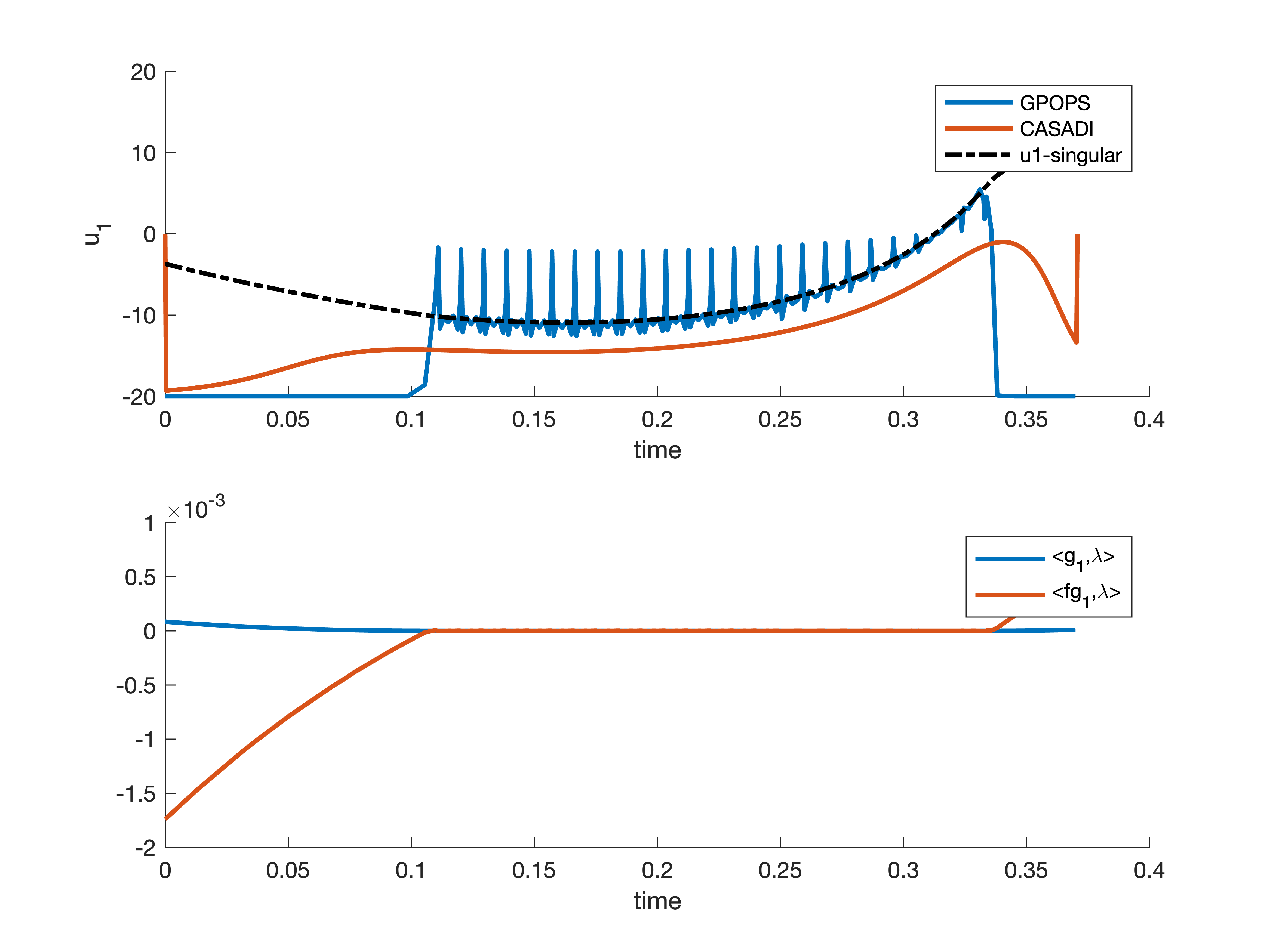}
            \caption{\small In this figure we illustrate how the results from this paper can be used to refine and obtain algebraic expressions for optimal control solutions that pass through singularity arcs. In the top plot we have the control signal for the first joint recovered by GPOPS-II (solid blue), CASADI (solid red), and the one predicted by \eqref{eq:uising_u1cont} using the costates from GPOPS-II (dashed black). On the bottom plot we have the values of $\phi_1=\ip{\lambda,g_1}$ (blue) and $\phi_1'=\ip{\lambda,fg_1}$ (red) computed by the costate values returned by GPOPS-II. Notice that when GPOPS-II converges to a singular solution for $\phi_1$, the recovered signal follows the same profile as the one predicted by theory, except for the numerical artifacts. Furthermore, the solution from CASADI struggles to identify the two distinct intervals.}
            \label{fig:GPOPS_satsingsat}
        \end{figure}

        The numerical artifacts present in both the GPOPS-II and the CASADI solution motivates the use of the theory presented in Section \ref{sec:uising} whenever singular arcs are detected, regularizing the intended control signal. To further exemplify this point, we input a different set of initial and final points in our numerical solvers for the 2DOF arm. Picking the initial and final points given by
        \begin{align*}
            x_0 = \begin{bmatrix}
                \frac{\pi}{20} \\ \frac{\pi}{20} \\ 0.5 \\ 0
            \end{bmatrix} \tag{\stepcounter{equation}\theequation}, ~~
            x_f = \begin{bmatrix}
                0.351541096001406 \\ 0.073883000198405 \\ 0.594756773574437 \\ -0.523743737608164
            \end{bmatrix},
        \end{align*}
        results in the signal for $u_1$ shown in Fig. \ref{fig:GPOPS_satsingsat}.

        By inspecting the top plot of Fig \ref{fig:GPOPS_satsingsat}, where the numerically recovered control signal is given in blue for GPOPS and red for CASADI, and the bottom plot, where the values of $\phi_1$ and $\phi_1'$ are given in blue and red respectively, we notice that for these initial and final points GPOPS-II recovers a solution with three clear subintervals: first being saturated on $u_1$, with $\phi_1>0$; second being singular; and third being saturated again with $\phi_1>0$. CASADI, however, struggles to find the distinct intervals of bang-bang and singularity, despite presenting basically the same final time. The signal in dashed black on the top plot, which is the value of \eqref{eq:uising_u1cont} for the costates given by GPOPS-II, recovers the numerical solution for the control signal in the singular interval without the artifacts.

\section{Conclusions}

    In this work, we review a theoretical approach to formulating singular extremals for the minimum-time control problem of fully actuated mechanical systems and illustrate its effectiveness on a 2-DOF robotic arm numerical example. We leverage and extend a previous result from one of the authors, which guarantees that not all control inputs of a given robotic system can be singular at the same time. We derived conditions for a singular extremal to be well defined and for an algebraic expression for the singular control signal to be obtainable. 
    
    For the case of a 2-DOF robotic arm, the theory simplifies greatly since if we impose, by choice of the input function, one of the joint controls to be singular in the sense of the Pontryagin Maximum Principle, we guarantee that the other will be bang-bang. For the numerical simulations of this paper, we impose that the first joint of our system be singular and the second be saturated and generate a trajectory that satisfies the maximum principle for a small enough interval of time. We also show that the alternate case (first joint bang-bang and second joint singular) is impossible according to the theory.
    
    Despite the Maximum Principle being only necessary for the optimality of a solution, the generated singular trajectory matches the collocation-based numerical solutions given by GPOPS-II and the shooting-based solution given by a CASADI-based algorithm built by the authors, while also giving an analytic expression for the control function and avoiding numerical artifacts. The matching between solutions is further confirmed when analyzing the recovered value for the switching function from the GPOPS-II solution, which is very close to zero but is incapable of being identically zero, instead oscillating with a very small amplitude.
    
    The method that we discuss is capable of dealing with trajectories that are singular and provides a closed-form expression for our singular control signals, allowing us to regularize the numerical solution whenever it enters a singular interval. While this paper explores mainly robotic arms and specifically the 2DOF arm, the properties that make this method viable still hold for a more complicated systems. By leveraging the more general theoretical framework, one could extend that for more interesting and complex systems.


\bibliographystyle{IEEEtran}
\bibliography{bibliography.bib}{}

%
\end{document}